\documentstyle[epsf]{MN}
\title[Modelling Nonlinear Gravitational Clustering]
{Modelling the Nonlinear Gravitational Clustering in the
Expanding Universe}
\author[T. Padmanabhan]
       {T. Padmanabhan \\
         Inter-University Centre for Astronomy and Astrophysics,
         Post Bag 4, Ganeshkhind, Pune 411007, India\\
         email:paddy@iucaa.ernet.in}
\begin{document}
\label{firstpage}
\maketitle

\def\cf{{\it cf.}\hskip 1.5pt}


\def\la{\mathrel{\mathchoice {\vcenter{\offinterlineskip\halign{\hfil
$\displaystyle##$\hfil\cr<\cr\sim\cr}}}
{\vcenter{\offinterlineskip\halign{\hfil$\textstyle##$\hfil\cr<\cr\sim\cr}}}
{\vcenter{\offinterlineskip\halign{\hfil$\scriptstyle##$\hfil\cr<\cr\sim\cr}}}
{\vcenter{\offinterlineskip\halign{\hfil$\scriptscriptstyle##$\hfil\cr<\cr\sim\cr}}}}}

\def\ga{\mathrel{\mathchoice {\vcenter{\offinterlineskip\halign{\hfil
$\displaystyle##$\hfil\cr>\cr\sim\cr}}}
{\vcenter{\offinterlineskip\halign{\hfil$\textstyle##$\hfil\cr>\cr\sim\cr}}}
{\vcenter{\offinterlineskip\halign{\hfil$\scriptstyle##$\hfil\cr>\cr\sim\cr}}}
{\vcenter{\offinterlineskip\halign{\hfil$\scriptscriptstyle##$\hfil\cr>\cr\sim\cr}}}}}

\begin{abstract}
 The gravitational clustering of collisionless particles in
an expanding universe is modelled using some simple physical ideas. I
show that it is indeed possible
to understand the nonlinear clustering in terms of three well defined
regimes: (1) linear regime (2) quasilinear
regime which is dominated by scale-invariant radial infall and (3)
nonlinear regime dominated by nonradial motions and mergers. Modelling
each of these regimes separately I show how the nonlinear two
point correlation function can be related to the linear correlation
function in heirarchical models. This analysis leads to results which
are in good agreement
with numerical simulations thereby providing an
explanation for numerical results. The ideas presented here will also
serve as a powerful anlytical tool to investigate nonlinear clustering
in different
models. Several implications of the result are discussed.
\end{abstract}

The driving force behind the formation of large scale structures in the
universe is the gravitational field produced by density fluctuations.
Overdense
regions accrete matter at the expense of underdense regions allowing
inhomogeneities in the universe to grow. Observations suggest
that the material content of the universe is dominated by dark matter,
likely to
be  made of collisionless elementary
particles. In that case, the gravitational force is mainly due to these
particles and, to first approximation, we can ignore the complications
arising from baryonic physics. The evolution of density perturbations is
then governed purely by the gravitational force.

When these density perturbations are small, it is possible to study
their evolution using linear theory. But once the
density contrast becomes comparable to unity, linear
perturbation theory breaks down and one must use
 N-body simulations to study the growth of perturbations.
While these simulations are of  some value in making concrete predictions
for specific models, they do not provide
clear physical  insight into the process of non-linear gravitational dynamics.
To obtain such an insight into  this complex problem,
it is necessary to model the
gravitational clustering of collisionless particles using simple
physical concepts. I shall develop one such model in this paper,
which - in spite of extreme simplicity - reproduces the simulation
results for hierarchical models fairly accurately. Further, this model
also provides insight into the clustering process and can be modified
to take into account more complicated situations.

The paradigm for understanding the clustering is based on the well
known behaviour of a spherically symmetric overdense region in the
universe. In the behaviour of such a region, one can identify
three different regimes of interest: (1) In the early stages of the
evolution, when the density contrast is small, the evolution is
described by linear theory. (2) Each of the spherical shells with
an initial radius $x_i$ can be parametersed by a mass contained inside
the shell, $M(x_i),$ and the energy, $E(x_i)$ for the particular
shell. Each shell will expand to a maximum radius $x_{max}\propto
M/|E|$ and then turn around and collapse. Such a spherical collapse
and resulting evolution allows a self similar description \cite{FILGOLD}
in which each shell acts as though it has an effective radius
proportional to $x_{max}$ \cite{BERT}. This will be the quasilinear phase. (3)
The spherical evolution will break down during the later stages due to
several reasons. First of all, non radial motions
will arise due to amplification of deviations from spherical symmetry.
Secondly, the existence of substructure will influence
the evolution in a non-spherically symmetric way. Finally, in the
real universe, there will be merging of such clusters [each of
which could have been centres of spherical overdense regions in
the begining] which will again destroy the spherical symmetry. This
will be the nonlinear phase.

The description given above is sufficiently vague and sufficiently
well known that one may suspect it can not lead to any insight into
the problem. In particular, real universe is hardly spherical. I
will show that it
is, however, possible to model the above process in a manner which
allows direct generalisation to the real universe.

To do this we will begin by studying the evolution of system
starting from a gaussian initial fluctuations with an initial power
spectrum, $P_{in}(k)$. The fourier transform of the power spectrum
defines the correlation function $\xi(a,x)$ where $a\propto t^{2/3}$
is the expansion
factor in a universe with $\Omega=1$. It is more convenient for our
purpose to work with the
average correlation function inside a sphere of radius $x$,
defined by
\begin{equation}
\bar{\xi}(a,x)\equiv {3\over x^3}\int^{x}_{0}\xi(a,y)y^2 dy\label{one}
\end{equation}
In the linear regime we have  $\bar{\xi}_L(a,x)\propto
a^2\bar\xi_{in}(a_i,x)$. In the quasilinear and nonlinear regimes,
we would like to have a prescription which relates the exact $\bar
\xi$ to the mean correlation function calculated from the linear
theory. One might have naively imagined that $\bar\xi(a,x)$ should
be related to $\bar\xi_{L}(a,x)$. But one can convince oneself that
the relationship is likely to be nonlocal by the following analysis:

 Recall that,
 the conservation of pairs
of particles, gives an exact equation satisfied by the
correlation function \cite{PEB1}:
\begin{equation}
{\partial\xi \over\partial t}+{1\over ax^2}{\partial\over\partial
x}[x^2(1+\xi)v]=0\label{qpaircorn}
\end{equation}
where $v(t,x)$ denotes the mean relative velocity of pairs at
separation $x$ and
epoch $t$. Using the mean correlation function $\bar\xi$ and a
dimensionless pair velocity $h(a,x) \equiv - (v/\dot{a}x)$, equation
(\ref{qpaircorn}) can be written as
\begin{equation}
({\partial\over\partial \ln a}-h{\partial\over\partial \ln x})\,\,\,
(1+\bar{\xi})=3h(1+\bar{\xi})
\end{equation}
This equation can be simplified by first introducing the variables
\begin{equation}
A=\ln a,\qquad X=\ln x ,\qquad D(X,A) = \ln (1+\bar{\xi})
\end{equation}
in terms of
which we have \cite{RNTP}
\begin{equation}
{\partial D\over\partial A}-h(A,X){\partial D\over\partial
X}= 3h(A,X)\label{qkey}
\end{equation}
 Introducing further a variable
$F=D+3X$, equation (\ref{qkey}) can be written  in a remarkably simple form as
\begin{equation}
{\partial F\over \partial A}-h(A,X){\partial F\over\partial
X}=0\label{qfunrel}
\end{equation}
The characteristic curves to this equation - on which $F$ is a
constant - are determined by
$(dX/dA)=-h(X,A)$ which can be integrated if $h$ is known. But note that
the charecteristics satisfy the condition
\begin{equation}
F=3X+ D=\ln [x^3(1+\bar{\xi})]={\rm constant}\label{four}
\end{equation}
or, equivalently,
\begin{equation}
x^3(1+\bar{\xi})=l^3\label{qxandl}
\end{equation}
where $l$ is another length scale. When the evolution is linear at all
the relevant scales, $\bar{\xi}\ll 1$ and $l\approx x$. As clustering
develops,
$\bar{\xi}$ increases and $x$ becomes considerable smaller than $l$.
It is clear that the behaviour of clustering at some scale $x$ is
determined by
the original {\it linear} power spectrum at the scale $l$ through the
``flow of information'' along the charesteristics.
This suggests that {\it we should actually
try to express the true
correlation function $\bar\xi(a,x)$ in terms of the linear correlation
function $\bar\xi_L(a,l)$ evaluated at a different point}.

Let us see how we can do this starting from  the quasilinear regime.
Consider a region surrounding a density peak in the linear stage,
around which we expect the clustering to take place. It is well known
that density profile around this peak
can be described by
\begin{equation}
\rho(x)\approx\rho_{bg}[1+\xi(x)]\label{five}
\end{equation}
Hence the initial mean density contrast scales with the initial shell
radius $l$ as $\bar\delta_i
(l)\propto\bar\xi_L(l)$ in the initial epoch, when linear theory
is valid. This shell will expand to a maximum radius of $x_{max}
\propto l/\bar\delta_i\propto l/\bar\xi_L(l)$. In  scale-invariant,
radial collapse, models each shell may be approximated as contributing with a
effective radius
which is propotional to $x_{max}$. Taking
the final effective radius $x$ as proportional to $x_{max}$, the final
mean correlation function will
be
\begin{equation}
\bar\xi_{QL}(x)\propto \rho\propto {M\over x^3}
\propto {l^3\over (l^3/\bar\xi_L(l))^3}\propto
\bar\xi_L(l)^3\label{six}
\end{equation}
That is, the final correlation function $\bar\xi_{QL}$ at $x$ is the cube of
initial correlation function at $l$ where $l^3\propto x^3
\bar\xi_L^3\propto x^3\bar\xi_{QL}(x).$ This is in the form demanded
by equation (\ref{qxandl})  if $\bar\xi\gg 1$. {\it Note that we did not assume
that
the initial power
spectrum is a power law to get this result.}

In case the initial power spectrum is a power law, with
$\bar\xi_{L}\propto x^{-(n+3)}$, then we  find that
\begin{equation}
\bar\xi_{QL}\propto x^{-3(n+3)/(n+4)}\label{qlndep}
\end{equation}
[If the
correlation function in linear theory has the powerlaw form $\bar\xi_{L}
\propto x^{-\alpha}$ then the process desribed above changes the index
from $\alpha$ to $3\alpha/(1+\alpha)$. We shall comment more
about this aspect at the end of the paper.]. For the power law case, the
same result can be obtained by more explicit means. For
example, in power law models the energy of spherical shell will scale
with its radius as some power which we write as
$E\propto x_i^{2-b}$. Since $M\propto x_i^3$, it follows that the
maximum radius reached by the shell scales as $x_{max}\propto
(M/E)\propto x_i^{1+b}$. Taking the effective radius as
$x=x_{eff}\propto x_i^{1+b}$,  the final density scales as
\begin{equation}
\rho\propto {M\over x^3}\propto {x_i^3\over x_i^{3(1+b)}}
\propto x_i^{-3b}\propto x^{-3b/(1+b)}\label{basres}
\end{equation}
In this quasilinear regime, $\bar\xi$ will scale like the density and we get
$\bar\xi_{QL}\propto x^{-3b/(1+b)}$.
The index $b$ can be related to
$n$ by assuming the the evolution starts at a moment when linear
theory is valid. The gravitational potential energy [or the kinetic
energy] scales as $E\propto x_i^{-(n+1)}$ in the linear theory. This
may be seen as follows: The power spectrum for velocity field,
$P_v(k)$ in
the linear regime is related to that of density by $P_v\propto
P(k)/k^2\propto k^{n-2}$. Hence the contribution to $v^2$ in
each logarithmic scale in k-space is $k^3P_v/2\pi^2\propto k^{n+1}
\propto x^{-(n+1)}$. Similarly, the gravitational potential energy
due to {\it fluctuations} is
\begin{equation}
\phi\propto \int_0^x 4\pi y^2 dy{\xi(y)\over y}\propto x^2\xi(x)
\propto x^{-(n+1)}
\end{equation}
So the total energy in the initial configuration scales as
$x_i^{-(n+1)}$ allowing us to determine $b=n+3$. This
shows that the correlation function in the quasilinear regime is the one given
by  equation (\ref{qlndep}) .

The case with power law initial spectrum has no intrisic scale, if $\Omega=1 $.
It follows that the evolution has to be self similar and
$\bar\xi$ can only depend on $q=xa^{-2/(n+3)}$. This allows us to
determine the $a$ dependence of $\bar\xi_{QL}$ by substituting $q$
for $x$ in equation (\ref{qlndep}). We find
\begin{equation}
\bar\xi_{QL}(a,x)\propto a^{6/(n+4)}x^{-3(n+3)/(n+4)}\label{qlax}
 \end{equation}
Direct algebra shows that
\begin{equation}
\bar\xi_{QL}(a,x)\propto [\bar\xi_{L}(a,l)]^3\label{qlscal}
\end{equation}
reconfirming the local dependence in $a$ and nonlocal dependence
in spatial coordinate.
This result has no trace of original assumptions [spherical evolution,
scale-invariant spectrum ....] left in it and hence once
would strongly suspect that it will have far general validity.

\begin{figure}
\epsfxsize=2.5truein\epsfbox[20 360 400 690]{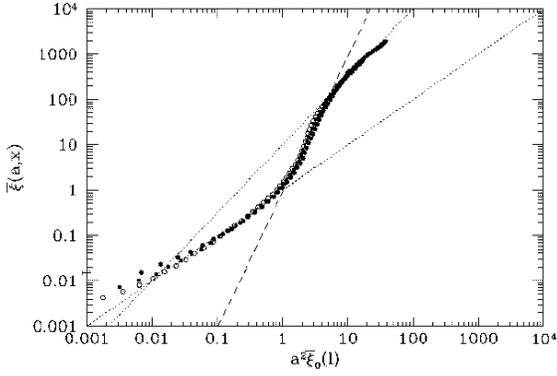}
\caption{Plot of $\bar\xi(a,x)$ against $\bar\xi_{L}(a,l)$ for CDM model.
The slopes in the three different regimes are indicated. The data points
are for three different redshifts [0.1,0.5 and 1.0] and are based on
the simulations described in Padmanabhan et al. (1995).}
\end{figure}

Let us now proceed to the third and nonlinear regime. If we ignore the
effect of mergers, then it seems reasonable that virialised systems
should maintain their densities and sizes in proper coordinates, i.e.
the clustering should be ``stable". This
would require the correlation function to have the form $\bar\xi_{NL}
(a,x)=a^3F(ax)$. [The factor $a^3$ arising from the decrease in
background density].
 From our previous analysis we expect this to be a function of
$\bar\xi_L(a,l)$ where $l^3\approx x^3\bar\xi_{NL}(a,x)$. Let us write
this relation as
\begin{equation}
\bar\xi_{NL}(a,x)=a^3F(ax)=U[\bar\xi_L(a,l)]\label{qtr}
\end{equation}
where $U[z]$ is an unknown function of its argument which needs
to be determined. Since linear correlation function evolves as
$a^2$ we know that we can write $\bar\xi_L(a,l)=a^2Q[l^3]$
where $Q$ is some known function of its argument. [We are using
$l^3$ rather than $l$ in defining this function just for future
convenience of notation]. In our case $l^3=x^3\bar\xi_{NL}(a,x)
=(ax)^3F(ax)=r^3F(r)$ where we have changed variables from
$(a,x)$ to $(a,r)$ with $r=ax$. Equation \ref{qtr} now reads
\begin{equation}
a^3F(r)=U[\bar\xi_L(a,l)]=U[a^2Q[l^3]]=U[a^2Q[r^3F(r)]]
\end{equation}
Consider this relation as a function of $a$ at constant $r$. Clearly
we need to satisfy $U[c_1 a^2]=c_2a^3$ where $c_1$
and $c_2$ are constants. Hence we must have
\begin{equation}
U[z]\propto z^{3/2}.
\end{equation}
Thus in the extreme nonlinear end we should have
\begin{equation}
\bar\xi_{NL}(a,x)\propto [\bar\xi_{L}(a,l)]^{3/2}\label{qnlscl}
\end{equation}
[Another way deriving this result is to note that if $\bar\xi=
a^3F(ax)$, then $h=1$. Integrating equation (\ref{qkey}) with appropriate
boundary
condition leads to  equation (\ref{qnlscl}).]
Once again we did not need to invoke the assumption that the
spectrum is a power law. If it {\it is } a power law, then we get,
\begin{equation}
\bar{\xi}_{NL}(a,x)\propto a^{(3-\gamma)}x^{-\gamma};\qquad
\gamma={3(n+3)\over (n+5)}
\end{equation}
This result is based on the assumption of ``stable clustering" and
was originally derived by Peebles \cite{PEB2}. It can be directly
verified that the right hand side of this equation can be expressed in
terms of $q$ alone, as we would have expected.

Putting all our results together, we find that the nonlinear mean
correlation function can be expressed in terms of the linear mean
correlation function by the relation:
\begin{eqnarray}
& \bar \xi_L (a,l) &\, ({\rm for}\,\, \bar \xi_L<1, \, \bar
\xi<1) \label{qtrial}\\
 \bar \xi (a,x)  = &{\bar \xi_L(a,l)}^3 & \, ({\rm for}\,\, 1<\bar \xi_L<5.85,
\, 1<\bar \xi<200)\nonumber\\
& 14.14 {\bar \xi_L(a,l)}^{3/2} & \, ({\rm for}\,\, 5.85<\bar\xi_L, \, 200<\bar
\xi)\nonumber
\end{eqnarray}
The numerical coefficients have been determined by continuity
arguments. We have assumed the linear result to be valid upto
$\bar\xi=1$ and the virialisation to occur at $\bar\xi\approx 200$
which is result arising from the spherical model.  The exact values of the
numerical coefficients can be  obtained only from simulations.

The true test of such a model, of course, is N-body simulations and
remarkably enough, simulations are very well represented by relations
of the above form.
Figure 1 shows the results of a CDM simulation based on the
investigations carried out in Padmanabhan et.al \shortcite{PCOS}. This data can
be fitted
by the relations \cite{BAGTP1} :
\begin{eqnarray}
 &\bar \xi_L(a,l) & \,({\rm for}\, \bar \xi_L< 1, \, \bar
\xi<1)\nonumber\\
\bar \xi(a,x)= &{\bar \xi_L(a,l)}^3 & \, ({\rm for}\, 1<\bar \xi_L<5, \, 1<\bar
\xi<125)\nonumber\\
&11.2 {\bar \xi_L(a,l)}^{3/2} &({\rm for}\, 5<\bar\xi_L, \, 125<\bar
\xi)\label{qbagh}
\end{eqnarray}
[The fact that numerical simulations show a correlation between
$\bar\xi(a,x)$ and $\bar\xi_L(a,l)$ was originally pointed out
by Hamilton et al. \shortcite{HKM} who, however, tried to give a multiparameter
fit to the data. This fit has somewhat obscured
the simple physical interpretation of the result though has the virtue
of being very accurate for numerical work.]

A comparison of equations (\ref{qtrial}) and (\ref{qbagh}) shows that the
physical processes
which operate at different scales are well represented by our model.
In other words, the processes descibed in the quasilinear and nonlinear
regimes for an {\it individual} lump still models the {\it average}
behaviour of
the universe in a statistical sense. It must be emphasised that the key
point is the ``flow of information" from $l$ to $x$ which is an exact
result.  Only when the results of the specific model are recast in
terms of suitably chosen variables, we get a relation which is of general
validity. It would have been, for example, incorrect to use spherical
model to obtain relation between linear and nonlinear densities at
the same location or to model the function $h$. With hindsight, it is
clear why such attempts have not succeeded in the past.

It may be noted that to obtain the result in the nonlinear regime,
we needed to invoke the assumption of stable clustering which has
not been deduced from any fundamental considerations. In case
mergers of structures are important, one would consider this
assumption to be suspect\cite{PCOS}. We can, however, generalise the above
argument in the following manner: If the virialised systems have
reached  stationarity in the statistical sense, the function $h$
- which is the ratio between two velocities - should reach some
constant value. In that case, one can integrate equation (\ref{qfunrel}) and
obatin the result $\bar\xi_{NL}=a^{3h}F(a^hx)$. A similar argument
will now show that
\begin{equation}
\bar\xi_{NL}(a,x)\propto [\bar\xi_{L}(a,l)]^{3h/2}\label{qnlscl}
\end{equation}
in the general case. For the power law spectra, one would get
\begin{equation}
\bar{\xi}(a,x)\propto a^{(3-\gamma)h}x^{-\gamma};
\gamma={3h(n+3)\over 2+h(n+3)}
\end{equation}
Simulations are not accurate enough to fix the value of $h$; in
particular, the asynptotic value of $h$ could depend on $n$
within the accuracy of the simulations. It may be possible to
determine this dependence by modelling mergers in some simplified form.

We conclude with two interesting speculations regarding the nonlinear
stage. If $h=1$ asymptotically, the correlation
function in the extreme nonlinear end depends on the linear index
$n$. One may feel that physics at highly nonlinear end should be
independent of the linear spectral index $n$. This will be the case
if the asymptotic value of $h$ satisfies the scaling
\begin{equation}
h={3c\over n+3}
\end{equation}
in the nonlinear end with some constant $c$. Only high ressolution
numerical simulations can test this
conjecture that $h(n+3)=$constant.

Also note that the radial, scale invariant infall described in
the quasilinear regime has the effect of changing the linear
correlation function $\bar\xi_L=x^{-(n+3)}=x^{-b}$ to the quasilinear
correlation function $\bar\xi_{QL}=x^{-3b/(1+b)}$. It is amusing to
ask what will be the effect of iterating this process N-times. It is
easy to see that the index after N iterations can be expressed in the form:
\begin{equation}
\gamma_N={A_N b\over 1+B_N b};\quad A_N=3^N;B_N={3^N-1\over 2}\label{qfxpt}
\end{equation}
The fixed point, of course, is $\gamma_{\infty}=2$ which is the only
nontrivial fixed point for such an evolution [with the other, trivial,
fixed point being zero]. If one could model the evolution as repeated
application of this process, one would expect a continuum of scaling
relations with the evolution being driven to a singular isothermal
sphere. The quasilinear evolution doesnot
change the $x^{-2}$ profile, a result which was noted earlier in Bagla and
Padmanabhan \shortcite{BAGTP2}. It is
not clear whether the clustering can indeed be modelled using equation
(\ref{qfxpt}).

The relations obtained in this paper will, of course, have certain limitations
on their validity. To begin with, we do expect a weak $n$-dependence
in these relations due to averaging over peaks of different heights. This
has been discussed using a simple analytic model, as well as numerically,
in T. Padmanabhan et.al,  \shortcite{PCOS} [Also see  Mo et.al.,
\shortcite{MJW} for a similar discussion]. Secondly, the asymptotic
behaviour will be sensitive to the value of $\Omega$. When $\Omega<1$,
structures ``freeze out" during the late stages of evolution and
``stable clustering" is likely to be a reasonable assumption. Finally,
models like HDM evolve in a manner very different from hierarchical models.
In the former, small scale power is generated by the breaking of long
wavelength
modes and the evolution is best modelled by instability of shell-like regions
in the
universe. Work is now in progress
to generalise the ideas of the present paper for other models.
\vspace{.20in}

\noindent {\bf Acknowledgements}
\vspace{.05in}

\noindent I thank J.S. Bagla, D. Lynden-Bell, R. Nityananda,
J.P. Ostriker and P.J.E. Peebles for several useful discussions. I thank
my coauthors  R. Cen, J.P. Ostriker
and F. Summers for permission to adapt a figure from our joint work. This work
was completed when I was visiting Institute of Astronomy, Cambridge in
Aug, 95 and I thank Ofer Lahav for hospitality.

\end{document}